\documentclass[preprint]{aastex}
\usepackage[super]{cite}
\usepackage{graphicx}
\usepackage{subfiles}

\begin{document}

\title{PRECISE ASTRONOMICAL FLUX CALIBRATION AND ITS IMPACT ON STUDYING THE NATURE OF THE DARK ENERGY}

\author{CHRISTOPHER W. STUBBS}

\affil{Department of Physics and Department of Astronomy, Harvard University, 17 Oxford Street\\
Cambridge, MA 02138, USA}
\email{stubbs@physics.harvard.edu}

\author{YORKE J. BROWN}

\affil{Department of Physics and Astronomy, Dartmouth College, 6127 Wilder Laboratory\\
Hanover, NH 03755, USA}
\email{yorke.brown@dartmouth.edu}



\begin{abstract}

Measurements of the luminosity of type Ia supernovae {\it vs.} redshift provided the original evidence for the accelerating expansion of the Universe and the existence of dark energy.  Despite substantial improvements in survey methodology, systematic uncertainty in flux calibration dominates the error budget for this technique, exceeding  both statistics and other systematic uncertainties.   Consequently, any further collection of type Ia supernova data will fail to refine the constraints on the nature of dark energy unless we also improve the state of the art in astronomical flux calibration to the order of 1\%.  We describe how these systematic errors arise from calibration of instrumental sensitivity, atmospheric transmission, and Galactic extinction, and discuss ongoing efforts to meet the 1\% precision challenge using white dwarf stars as celestial standards, exquisitely calibrated detectors as fundamental metrologic standards, and real-time atmospheric monitoring.

\end{abstract}


\section{Introduction}
\label{intro}

Before the development of astronomical CCDs, photoelectric detectors routinely attained photometric precision at the parts-per-thousand level.\cite{Young}  The advent of CCD cameras, by enabling simultaneous photometric measurement across the entire focal plane, opened up numerous fruitful avenues of research, notably the massive surveys of type Ia supernovae that resulted in the discovery of the cosmic acceleration.\cite{Highz,SCP}.  For a number of reasons, however, this new convenience and efficiency came at a cost in precision, with a typical CCD-based photometric survey achieving precision of only about 3-5\%.  Such a  level of precision was entirely adequate for the discovery of the dark energy---a 20\% effect---but current and future attempts to use supernovae to constrain the nature of the dark energy are placing increasingly stringent demands on the calibrations that underpin these measurements \cite{DES,LSST,WFIRST,EUCLID}  Current and planned surveys produce sufficient numbers of supernovae (and photons per supernova) that statistics no longer dominate the photometric error budget; it is now systematic uncertainties in basic flux calibration that limit the utility of survey data.\cite{SNLS1,SNLS2,Conley2011,PanSTARRSsystematics}  In order to move forward, CCD-based photometric technique must reclaim the precision once provided by its more primitive photoelectric ancestors.

This paper outlines the nature of the precision photometry challenge in the context of type Ia supernova cosmology and describes a path to the sub-1\% precision regime through the use of a flux calibration approach that departs from the tradition of using celestial standards and embraces the use of well-characterized detectors as the fundamental metrologic standard.

\section{Scientific Motivation}
\label{motivation}

\subsection{Mapping the Cosmic Expansion Using Type Ia Supernovae}

\begin{figure}[tph!] 
\centerline{\includegraphics[width=6in]{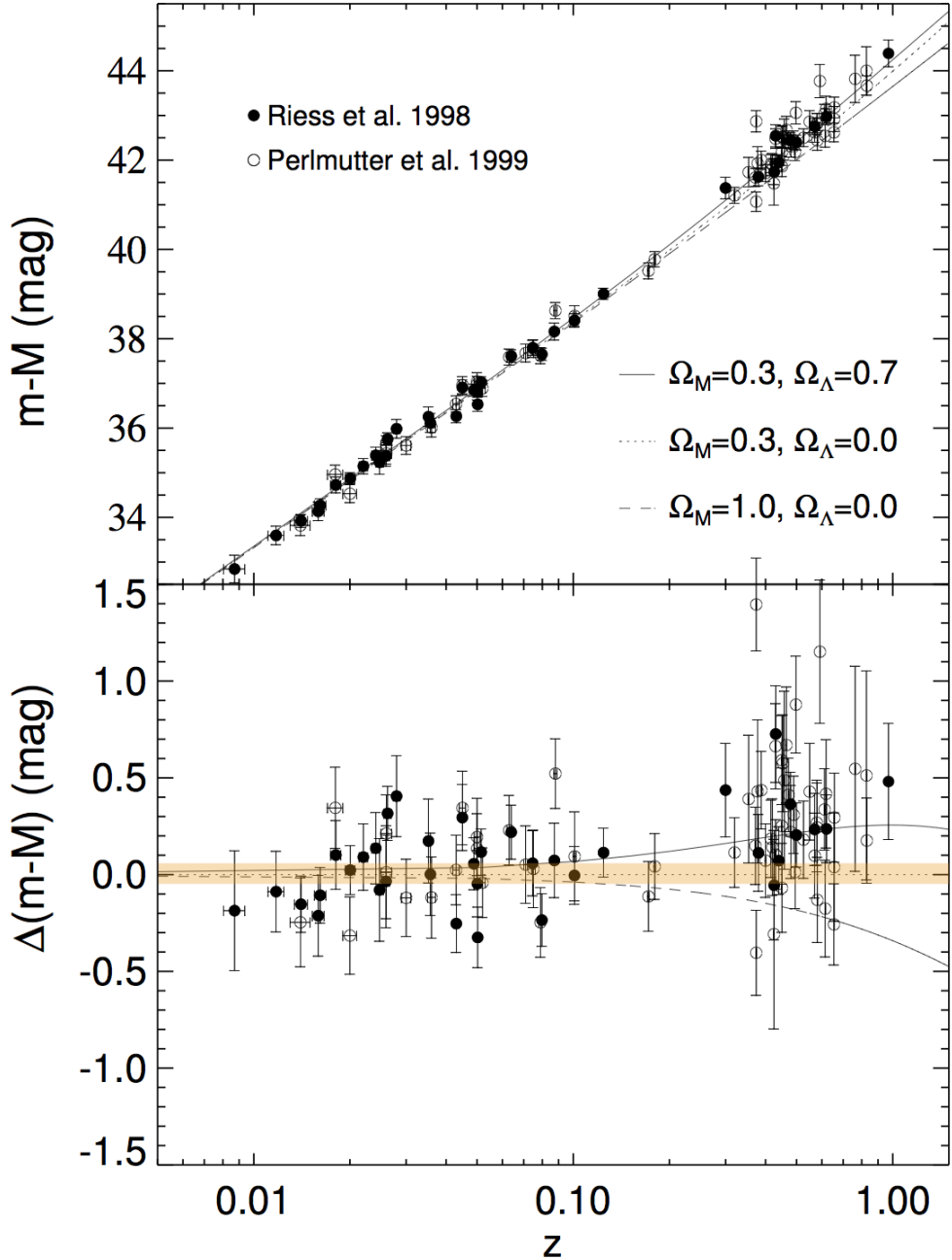}}
\vspace*{3pt}
\caption{Type Ia supernova Hubble diagram, showing luminosity distance {\it vs.} redshift.  The data points shown are from the original supernova surveys\cite{Highz,SCP} that established the cosmic acceleration.  The supernovae at higher redshift are $\sim20\%$ fainter than would be expected in a matter-dominated critical density Universe (the dashed curve). The orange bar corresponds to a 1\% flux uncertainty, which is the calibration target for next-generation measurements to constrain the nature of dark energy. (From Riess\cite{Riess2000}, used with permission).
\protect
\label{fig:Hubble}}
\end{figure}

The initial evidence for the accelerating expansion of the Universe was obtained from measurements of the brightness of type Ia supernovae as a function of redshift.\cite{Highz,SCP} The peak brightness of these cosmic detonations (after an empirical correction for dependence on color and rate-of-decline of the light curve) serves as a standard candle and can therefore be converted into a distance measure called the ``luminosity distance."  The cosmological redshift, which is obtained from the wavelength shift of spectral features, provides a measure of cosmic expansion during the light travel time.   Combining the observed peak brightness of a supernova with its cosmological redshift provides a direct measurement of the amount of cosmic expansion since the time the light was emitted by the supernova.  This relationship is often presented as a graph of luminosity distance {\it vs.} redshift called the ``Hubble diagram"---a graphical representation of the expansion history of the universe.  Distant supernovae (at redshift z$\sim$0.8) are roughly 20\% fainter than one would expect in a critical density, matter dominated ($\Omega_{matter}$=1), Universe (where $\Omega$ denotes dimensionless cosmic density normalized to the critical density).  Thus, the cosmic expansion is seen to be accelerating---a phenomenon presumed to be due to a still-mysterious dark energy.  The initial supernova datasets that indicated a non-zero dark energy content are shown in Figure \ref{fig:Hubble}.

As the study of the dark energy progresses from the discovery phase to the detailed characterization phase, we face the challenge of making increasingly precise measurements of the expansion history.

It is common to parameterize the relationship between the density $\rho_i$ and pressure $P_i$ of the various constituents of the Universe (matter, radiation, dark energy, and so forth) in terms of an equation of state parameter $w_i$, where $w_i=P_i/\rho_i$. Measuring the equation of state parameter $w$ of the dark energy, and any redshift-dependence $w=w(z)$, is a primary objective in contemporary fundamental physics.  Physically, this measurement amounts to determining whether the dark energy (or vacuum energy) density is invariant over cosmic time, in which case $w(z)$ would be constant.  

Ascertaining whether there is any significant evidence for $w\ne -1$ or for $w=w(z)$ is critical to discriminating between an interpretation of dark energy as a manifestation of the cosmological constant originally envisioned by Einstein, or whether we should instead attribute this effect to some other unexpected aspect of fundamental physics. The stakes are high from the perspective of fundamental physics, and our ability to continue to use type Ia supernovae as a probe of the nature of dark energy depends on our ability to drive down the dominant calibration contribution to the systematic error budget.

For supernova surveys that use spectroscopic measurements to determine redshifts, the uncertainty in redshift is overwhelmed by the uncertainty in luminosity distance---which is based on flux measurements.  Relative flux calibration as a function of wavelength is presently the dominant source of uncertainty for using type Ia supernovae as a probe of the history of cosmic expansion.  In order to distinguish among different interpretations of the dark energy we will need relative flux measurements at a much improved level of precision.

\begin{figure}[ph]
\centerline{\includegraphics[trim={0in 2in 0in 2in},clip, width=\textwidth]{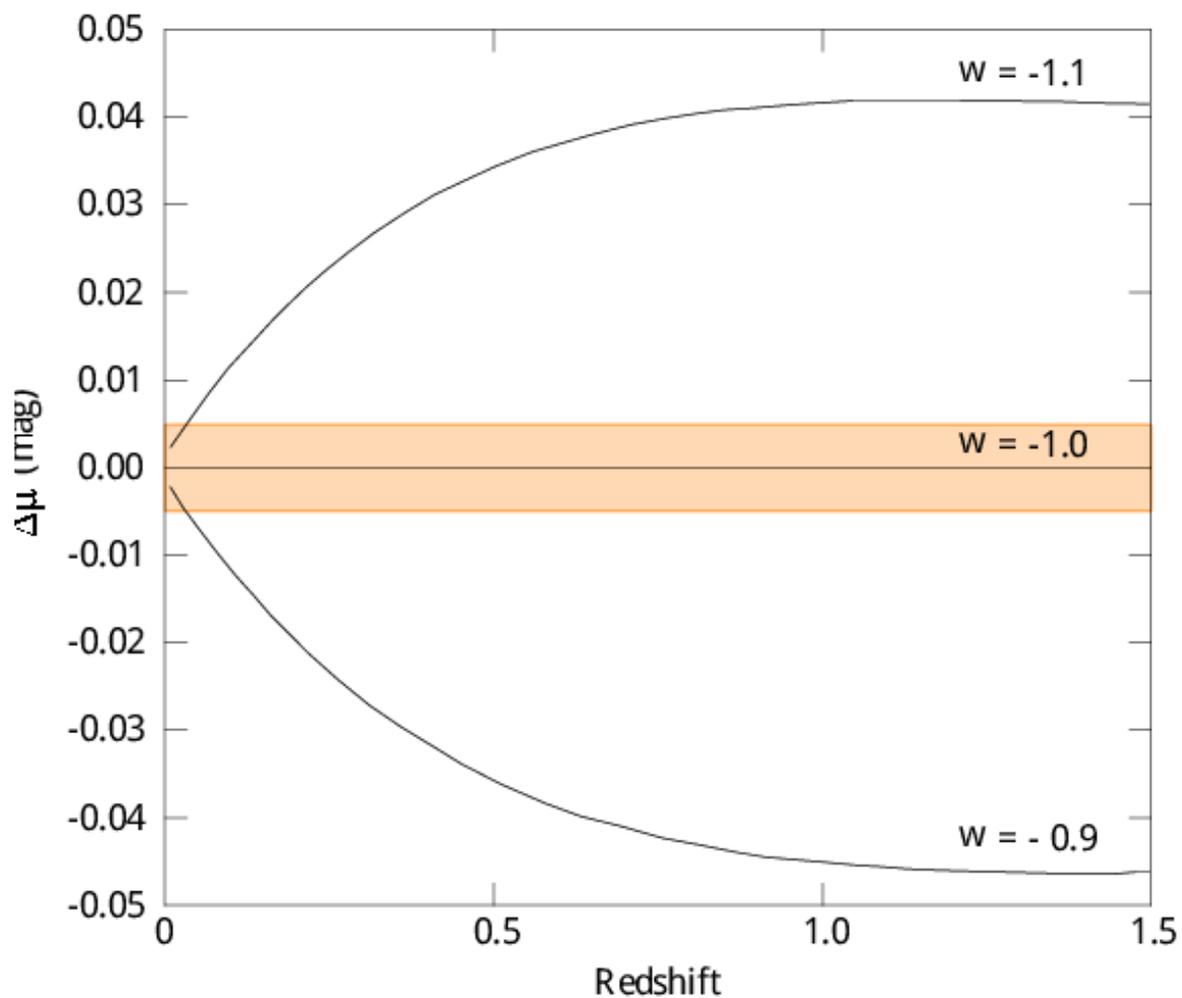}}
\vspace*{8pt}
\caption{Dependence of luminosity distance, in magnitudes, on the equation of state parameter $w$, \textit{vs.} redshift. Contemporary observations constrain the equation of state parameter to lie within 10\% of $w=-1$ and so achieving substantially better constraints requires sub-percent precision in flux measurements. A magnitude difference of 0.01 corresponds to a 1\% flux change.  The orange bar represents an uncertainty of 1\%.
\protect
\label{fig:wplot}}
\end{figure}

Figure \ref{fig:wplot} shows how the luminosity distance \textit{vs.} redshift depends on the value of the equation of state parameter $w$.  This plot assumes $\Omega_m = 0.3$ and zero curvature. Current data constrain the equation of state parameter to be within 10\% of $w=-1$, and producing substantially more refined constraints will require flux measurements with precision of better than 1\%. 

Table \ref{tab:contrib} shows the contributions of various sources of uncertainty to measurements of $w$ from the SNLS survey. Notice that calibration systematics completely dominate this error budget, including contributions from uncertainties in the flux standard BD+17 and in the instrumental sensitivity function.  

\begin{table}[h]
\centering
\caption{Contribution of various sources to the uncertainty of $w$, the equation of state parameter of the dark energy, for type Ia supernovae.  These data are taken from Sullivan {\it et al.}\cite{Sullivan2011}}
{\begin{tabular}{@{}lcc@{}} 
\hline
Uncertainty sources & $\sigma_w$ & $\sigma$/$\sigma_{statistical}$ \\
\hline
All Systematics & 0.080 & 1.47 \\
Calibration systematics only & 0.074 & 1.37 \\
~~~~Colors of BD+17 & 0.064 & 1.18 \\
~~~~SED of BD+17 & 0.062 & 1.15 \\
~~~~SNLS zero points & 0.060 & 1.11 \\
~~~~SNLS filters & 0.058 & 1.08 \\
\hline
Statistical & 0.054 & 1.00 \\
\hline
\end{tabular}
\label{tab:contrib} }
\end{table}

\subsection{More Refined Questions Require Improved Precision}

The supernova evidence for dark energy was obtained from $\sim$20 objects, with a photometric precision of approximately 10\% per object, which corresponds to 0.1 
astronomical magnitudes. Table \ref{tab:numbers} shows how the number of well-measured type Ia supernovae has been increasing by roughly a factor of ten with each successive generation of supernova cosmology survey, and the improvement in flux calibration needed to capitalize on the anticipated $1/\sqrt{N}$ reduction in statistical uncertainty. 

\begin{table}[h]
\centering
\caption{Supernova Surveys, Number of Objects, and Requisite Flux Calibration Uncertainty for $\sqrt{N_{\textnormal{SNe}}}$ scaling in statistical uncertainty.}
{\begin{tabular}{@{}rrc@{}}
\hline
Data Set & Number of SNe & Systematic Uncertainty \\
~ & ~ & in Flux Calibration\\
\hline
Discovery Data (1998) & $\sim$ 20 & 10\%\\
ESSENCE, SNLS (2005) & $\sim$ 300 & 3\% \\
PanSTARRS, DES (2015) & $\sim$ 2000 & 1\% \\
LSST (2020) & $\sim$ 20,000 & 0.3\% \\
\hline
\end{tabular}\label{tab:numbers} }
\end{table}

Current supernova surveys have the potential to discriminate among proposed theories of the dark energy, but only if flux calibration follows the trend in statistical uncertainty into the sub-percent regime.  Indeed, without attention to systematic uncertainties in flux measurement, newer surveys offer little improvement in precision over their antecedents.

\section{Statement of the Problem}
\label{problem}

Luminosity distance is a measure of the distance to a source as computed by a naive application of the inverse square law.  If a source of luminosity $L$ produces a received flux $\Phi$ at the detector, that flux obeys the inverse square law
\begin{equation}\label{flux}
\Phi = \frac{L}{4 \pi D_L^2}A ~,
\end{equation}
where $D_L$ is the luminosity distance to the source and $A$ is the collecting area of an ideal instrument.  Thus the luminosity distance is given by 
\begin{equation}\label{distance}
D_L = \sqrt{\frac{AL}{4 \pi \Phi}} ~.
\end{equation}
The luminosity distance therefore depends only on the \emph{ratio} of the intrinsic luminosity of the supernova to the flux captured by the observing instrument, that is, to the \emph{difference} between the absolute magnitude $M$ and observed magnitude $m$.  Furthermore, the specific value of the luminosity distance is not necessary for our purposes; we only need to know the ratios of luminosity distances at different redshifts in order to infer the history of the cosmic expansion.  Consequently neither the absolute intrinsic luminosity nor the collecting area and other efficiencies are needed; just the ratio of $L/\Phi(z)$, or equivalently, the difference $m(z)- M$.  This difference is often called the ``distance modulus" $\mu \equiv m - M$.

This definition of luminosity distance assumes two things: first that the detector captures \emph{all} the flux falling on it; and second, that the light is not affected by traversing the intervening space from source to detector.  Neither of these assumptions is valid.  As the light from a supernova travels to an observer's instrument, it is partially extinguished by several wavelength-dependent media---including the bandpass filters and detectors that are a part of the instrument itself.  Different supernovae at different redshifts interact with these media differently, due to their redshifted photon spectra.  A definition of luminosity distance valid across the entire range of relevant redshifts demands that all scattering, absorption, and stray light effects be taken into account. In particular, it is vital that we establish the wavelength-dependence of the instrumental sensitivity function. 

If multiple data sets, using different surveys and telescopes in order to span a wider range in redshift, are combined into a joint analysis then the overall flux normalization of the various surveys is an important calibration concern\cite{ScolnicCalib}. However we will assume a homogeneous data set, acquired with a single instrument and telescope, so that the primary instrumental calibration issue is the instrumental sensitivity function {\it vs.} wavelength.  

The problem of characterizing the dark energy straddles the boundary between physics and astronomy.  In order to promote understanding between physicists and astronomers, we offer a description of the problem in each of the languages of these two disciplines.

\subsection{The Problem Stated in the Language of Experimental Physics}

Imagine that all type Ia supernovae emitted light at a single wavelength. Measuring the apparent brightness of a collection of supernovae across a range of distances entails a comparison of detected flux at different observer-frame wavelengths, since the light from more distant objects arrives at longer wavelengths due to cosmological redshift. The basic measurement problem is to distinguish measurement artifacts or calibration errors in instrumental photon detection sensitivity {\it vs.} wavelength from the cosmological signatures of interest. What matters here is the \emph{precision} of the relative instrumental photon sensitivity calibration {\it vs.} wavelength. The supernova-based measurement of the Hubble diagram is a relative measurement of the flux ratio obtained from sources at different redshifts. The actual peak luminosity at some reference redshift, in units of joules per nm per sec per unit collecting area is irrelevant, and is bundled into an overall common multiplicative nuisance parameter that includes the telescope collecting area, the intrinsic peak brightness of a type Ia supernova, and the Hubble expansion rate at the present time.  If this claim is true for a $\delta$-function emitted photon spectrum, then it's also true for any spectrum of emitted supernova photons. The instrumental calibration goal is to fully characterize the terms that collectively determine the number of photons detected in an instrument. For ground-based instruments, this characterization includes the attenuation and scattering that occurs in the Earth's atmosphere.

The supernova flux measurements are typically obtained in a series of broadband filters with typical optical bandwidths of 100 nm.  For a particular filter labeled $j$, the number of detected supernova photons $N^\gamma_{d}$ is the integral over wavelength of the emitted photon spectrum times the various transmission factors, times the instrumental sensitivity function.  Thus for a measurement through filter $j$, cast in terms of the observer frame wavelength $\lambda$, 
\begin{equation}\label{throughput}
N^\gamma_{d}(j)= K \int S(\lambda)~N(\lambda,z)~G(\lambda, z)~A(\lambda)~R(\lambda, j)~C(\lambda,z,\Omega_m,\Omega_\Lambda,w...) ~d\lambda ~, 
\end{equation}
\noindent
where $K$ is an overall normalization that includes factors such as the collecting area and intrinsic supernova brightness, $S$ is the source photon spectrum in the observer frame, $N$ is the near-object transmission (due to scattering from dust in the supernova's host galaxy, for example), $G$ is Galactic transmission through the dust layers in the Milky Way, $A$ is the optical transmission through the Earth's atmosphere (which acts as a time-variable optical filter that we need to correct for), and $R(\lambda, j)$ is the instrumental response function, {\it i.e.} the wavelength-dependence of the photon detection efficiency of the apparatus, excluding the atmosphere but including the detector's quantum efficiency and the transmission of broadband filter $j$, and $C(\lambda,z)$ is the cosmological transfer function that connects the emitted photon spectrum at a given redshift and emission wavelength into the observer-frame photon spectrum.  It is this function $C(\lambda,z,\Omega_m,\Omega_\Lambda,w \dots)$ that contains the information of primary interest to physicists.  

Our measurement goal is to infer $C(\lambda, z, \Omega_m,\Omega_\Lambda, w \dots)$ from measurements of $N^\gamma_{d}$ in the different filters used in the photometric survey. This goal requires a determination of the various sources of line-of-sight attenuation, including the region around the supernova, wavelength-dependent scattering from particulate matter in the Milky Way, atmospheric attenuation (for ground-based measurements) as well as a means for calibrating the relative instrumental sensitivity function across the different filter passbands used.  

Each of the terms in equation \ref{throughput} merits careful and thorough attention. The wavelength-dependence of the extra-Galactic and Galactic photon scattering is typically used to measure or constrain the $N$ and $G$ terms, and combined optical and infrared measurements have recently played an important role.\cite{IRSN}  Independent estimates of Milky Way extinction can be obtained both from precise stellar photometry and infrared emission from Milky Way dust.\cite{SFD,Green}. Transmission of light through the Earth's atmosphere can be measured directly, with independent apparatus.\cite{Stubbs2007,LSST,DES}  In this paper we will focus our discussion on the determination of the instrumental response function.  

The determination of the instrumental photon response function $R(\lambda,j)$ unavoidably requires calibrating the instrument against some known photon spectral flux standard. In order for supernovae to remain a competitive probe of the nature of the dark energy we will need to accomplish this calibration to the sub-percent level.

\subsection{The Problem Stated in the Language of Astronomical Photometry}

Translating the calibration considerations described above into the language of astronomical photometry, the overall zeropoint of the photometric system is cosmologically uninteresting, and can be included in the nuisance parameter that includes the absolute magnitude of a type Ia supernova and $h_{100}=H_0/100$ km/sec/Mpc. The photon flux calibration that is vital for type Ia supernova cosmology is the conversion from instrumental magnitudes to incident photon arrival rates, across the different passbands in the system. We need to know, at the sub-percent level for LSST and WFIRST, how to connect instrumental magnitude differences between, say, the $r$ and $i$ bands to the actual incident photon ratio at the top of the atmosphere. This objective requires a calibration that allows us to interpret a measured $(r-i)$ color of an object as a known ratio of photon arrival rates integrated across the two passbands. 

This calibration cannot be accomplished by mapping the natural photometric system of the supernova survey onto a pre-existing standard empirical photometric system, or vice versa, since the existing ``standard photometric systems'' are themselves based on some (often historical) photon flux calibration. We stress that the goal is sub-percent precision ({\it i.e.} a few millimagnitude) in our knowledge of the difference in photometric zeropoints between passbands. 

We can assess the prospect of achieving this desired performance by comparing to the precision achieved in various contemporary astronomical photometry campaigns. The all-sky uniformity of the photometry from the SDSS in a single passband is reported to be at the 2\% level.\cite{SDSS} For LSST, the band-to-band calibration uncertainty onto the AB magnitude system for level-2 processed data is 5 mmag or 0.05\%.\cite{LSST}

\begin{figure}[ph]
\centerline{\includegraphics[width=\textwidth]{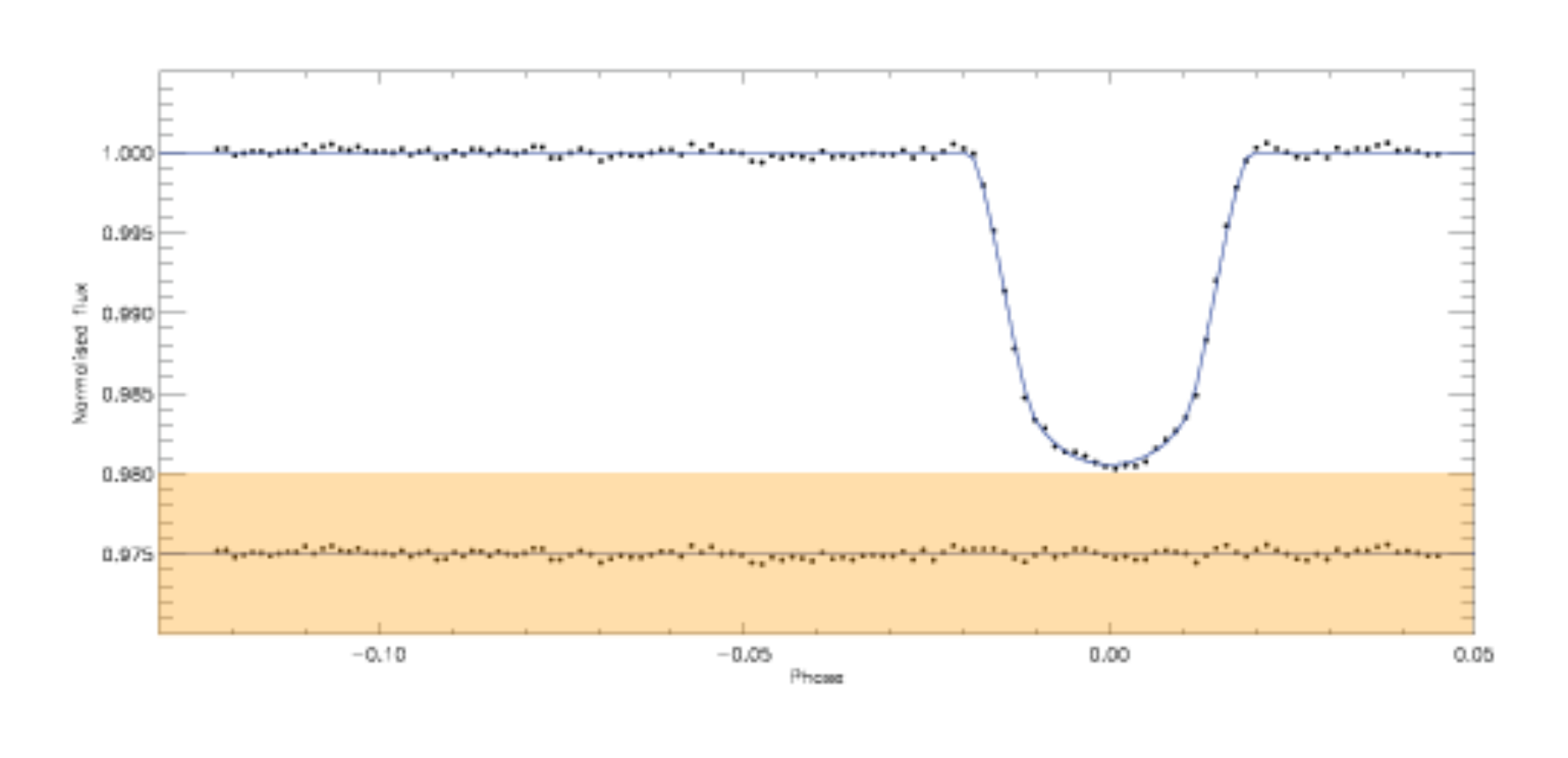}}
\vspace*{8pt}
\caption{Single-band high-precision differential photometry is readily achievable from the ground.  The orange bar corresponds to a 1\% flux uncertainty.  (From J. Tregloan-Reed \& J. Southworth \cite{Tregloan2013}, used with permission).
\protect
\label{fig:diff-photo}}
\end{figure}

Exoplanet searches have motivated the exquisitely precise measurements made possible by \emph{differential} photometry.  For example, Kepler achieves photometric precision at the 100 ppm level over 6-hour periods \cite{Kepler}.
Figure \ref{fig:diff-photo} shows an example of ground-based differential photometry that has achieved 0.2 mmag rms precision.\cite{Tregloan2013}

Clearly CCDs and telescopes are able to make single-band photometric measurements at the requisite level of precision. But the routine methods of broadband photometry (simple secant(zenith) extinction corrections, for example) do not support this level of precision. Monitoring temporal and azimuthal variability in atmospheric transmission, and proper corrections that include the interplay between atmospheric extinction and the source spectrum ({\it i.e.} color-airmass terms), will be essential to achieving the desired level of precision. With sufficient attention to detail and with a keen ongoing campaign to identify and suppress sources of systematic errors in photometry, we can hope to attain the desired few-millimagnitude single-band precision---although the question of how to precisely calibrate measurements made \emph{between} passbands remains an open issue. 

What spectrophotometric standard or combination of celestial standards can support photometric calibration at this level? We explore this issue in the section that follows. 

\section{Options for Relative Astronomical Photon Flux Calibration}
\label{standards}

We seek a calibration standard that can be used to determine the relative photon response sensitivity function of an astronomical imaging system, across the different filters of the survey.  There are three conceptually distinct calibration alternatives, each using a different physical principle as the basis for flux calibration across different wavelengths: 

\begin{enumerate}
\item A blackbody source at a known temperature and emissivity, in conjunction with our theoretical understanding of blackbody emitters and the Planck emission spectrum, can be used to generate a known photon spectrum. If this calibration source is on the ground, this knowledge needs to be transferred onto the sky somehow, typically by using stars as transfer standards and pointing an instrument alternately at the blackbody calibration source and at celestial sources. 

\item Theoretical knowledge of the physics of hydrogen can be used to derive the photon emission spectrum of DA white dwarf stars as a function of two parameters: effective temperature and surface gravity. 

\item The SI definition of the electrical watt can be linked to measurements of optical power using the method of electrical substitution radiometry. This method is 
used\cite{NIST1621} by NIST to calibrate Si photodiodes as transfer standards, for which they provide quantum efficiency \textit{vs.} wavelength measurements tied to their implementation of the SI unit of electromagnetic radiant intensity. 
\end{enumerate}

In the first two cases listed above, our understanding of the basic physics is used in conjunction with an emissive source (at the wavelengths of interest) to produce a ``known'' spectrum. In the third case the metrology standard is a detector of known quantum efficiency, rather than an emissive source. To our knowledge all optical and infrared astronomical relative spectral calibrations are based on one of these three alternatives, albeit often though a chain of multiple transfer standards. 

Before exploring the implementation of these different methods, we note some of the perils of using fluxes from \emph{any} celestial object as the foundational element for relative flux calibration. The first concern is the assumption of temporal stability in the emitted flux. Sources of temporal variability abound, including star spots, eclipsing of binary systems, and intrinsic instabilities in stellar atmospheres. The second concern, pertinent specifically to DA white dwarfs, is extinction by intervening material. The theoretical spectrum computed from modeling of the stellar atmosphere is not necessarily found in the flux we receive here at Earth. Interstellar extinction can corrupt the photon spectrum arriving at the top of the atmosphere, effectively introducing free parameters into what is supposed to be a definitive model.

\subsection{Terrestrial Blackbody Sources}

The star Vega, which has served as the calibration basis for a number of astronomical photometric campaigns, has served the community well for photometry at the 5-10\% level, but the photon spectrum of this dust-enshrouded, pole-on rotating star on the verge on centripetal breakup is not known \textit{a priori}. Rather, it serves as a transfer standard. Hayes and Latham\cite{Hayes1975a,Hayes1975b} used astronomical spectrophotometric instruments to observe the spectra of terrestrial blackbody sources, and alternated between observations of Vega and these terrestrial standards. Vega serves as a zero-magnitude transfer standard, transporting our knowledge of blackbody spectra into the celestial domain. But it's far too bright to observe directly with typical astronomical instruments and telescopes, and so we resort to either fainter ``spectrophotometric standards'' with photon spectra that are now two steps removed from the actual calibrated terrestrial source,  or ``photometric standards'' whose broadband magnitudes are tied to the broadband magnitudes of Vega in a given set of photometric filters. Differences in system response functions are mapped out as a function of stellar color, to derive ``transformation equations'' to transfer between a survey's natural photometric system and a Vega-based ``standard system."  Systematic uncertainty in the monochromatic (555.7 nm) flux of Vega is assessed as being
0.5\%.\cite{BohlinVega} We note here some excerpts from the foundational paper in which Hayes and Latham\cite{Hayes1975b} established Vega as flux standard.  The first excerpt pertains to the exponent in the Angstrom approximation for attenuation by atmospheric aerosols, where the cross section for aerosol attenuation scales as $(\lambda/\lambda_0)^{-\alpha}$:
 
\begin{quote}``We cannot account for why the solar observations at Mount Lemmon and Mount Hopkins give a much larger value of $\alpha$, and we have arbitrarily adopted $\alpha=0.8$ as appropriate for nighttime photometric conditions. This value is smaller than those quoted in the literature on atmospheric aerosols, which usually refer to lower altitudes and poorer transparency \dots''
\end{quote}

\noindent
The second excerpt pertains to the sources used, which were vats of molten metal placed on California mountaintops:

\begin{quote}
``For the platinum blackbody, Oke and Schild adopted a temperature 6 K below the standard freezing point of platinum in order to get agreement with their other sources, thus destroying the value of the platinum blackbody as a fundamental source. Furthermore, they quote an uncertainty of 5 percent, which is two to three times worse than the uncertainties they quote for the lamp and copper blackbody results. Because of these two problems, we have rejected all their platinum blackbody data \dots''
\end{quote}

We must emphasize here that we do not intend any criticism of the heroic efforts made by these pioneers of astrophotometric calibration, nor of the extensive program of Vega-based photometric standards that have been established by Landolt\cite{Landolt1992}.  Nevertheless, we must argue that calibration methodologies incorporating Vega as a standard simply cannot support sub-percent, cross-band calibration.
 
The most recent attempts to use terrestrial blackbody sources as the foundation for astronomical flux calibration were in the previous century, and we judge this method as having fundamental limitations that make it difficult to use for sub-percent flux calibration. Challenges include the difficulty in knowing wavelength-dependent emissivity, establishing the triple point temperature of vats of molten metal, and correcting for both the horizontal and vertical components of atmospheric attenuation. In our assessment this path will not prove fruitful in achieving the calibration objectives demanded by current and future supernova surveys. 

\subsection{DA White Dwarf Stars and First-Principles Modeling of Hydrogen Atmospheres}

\begin{figure}[ph]
\centerline{\includegraphics[width=\textwidth]{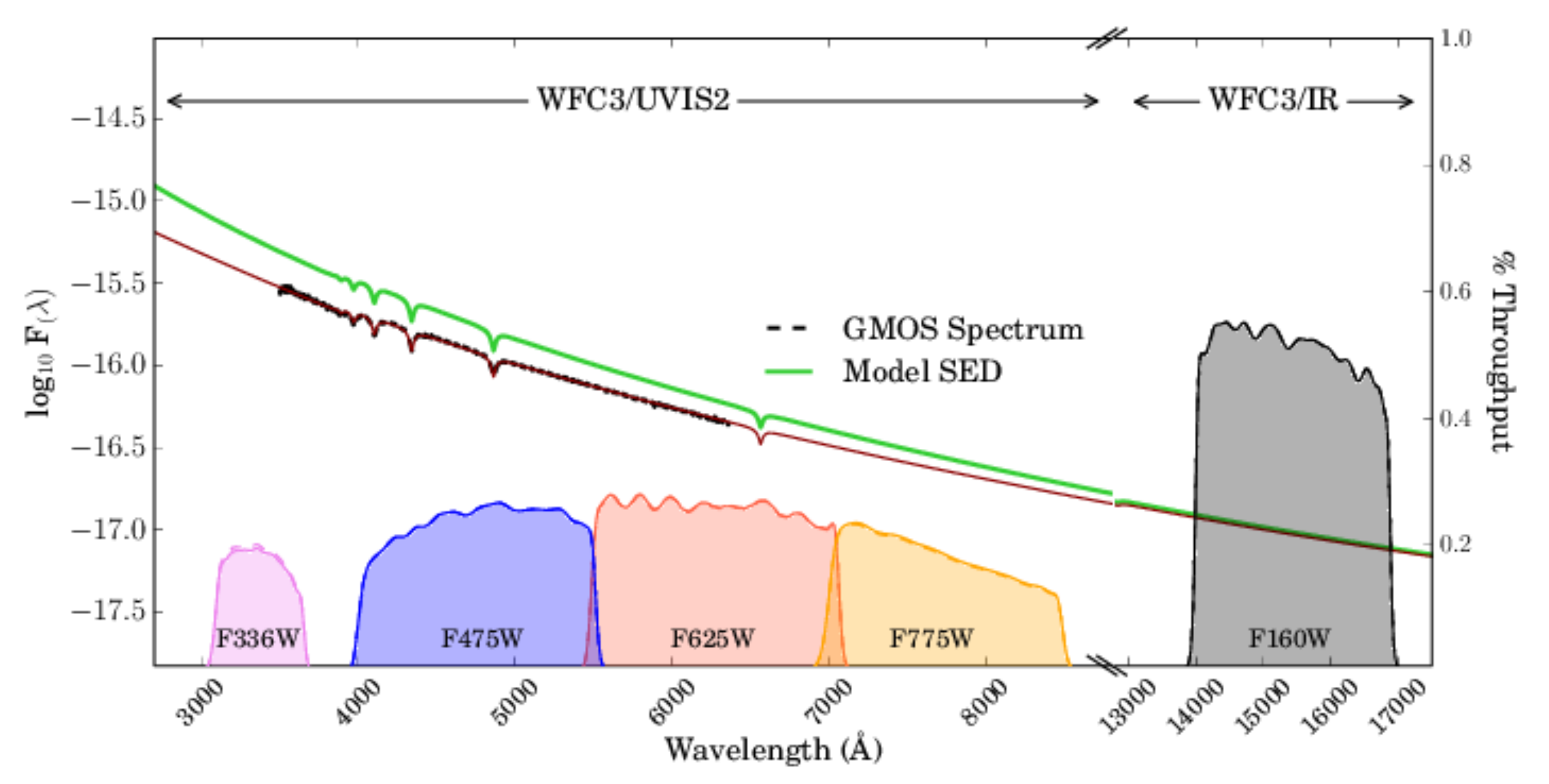}}
\vspace*{8pt}
\caption{Broadband measurements of faint DA white dwarf stars with HST can provide calibrators that are within the dynamic range of HST. (From G. Narayan {\it et al.}.\cite{Narayan2015}, used with permission.) A combination of space-based photometry at optical and NIR wavelengths can be combined with ground-based spectra to constrain both Galactic extinction and model atmosphere parameters. The figure shows the throughputs of the HST passbands used, the ground-based spectrum (black points) obtained at Gemini that was used to establish $\log(g)$ and effective temperature, the model spectrum with (red line) and without (green line) Galactic extinction.  
\protect
\label{fig:HST_WD}}
\end{figure}

A DA white dwarf star with a temperature between 20,000 and 80,000K is a remarkably simple system---a gravitationally bound, burnt-out stellar remnant with a hydrogen atmosphere. Its emitted photon spectrum should therefore be a function of two parameters that can be measured with ground-based spectroscopy: the effective temperature $T_{\textnormal{eff}}$ and the surface gravity, $\log(g)$.  Using just these parameters, we should be able to invoke our understanding of Hydrogen and radiation transfer to construct a first-principles model that accurately predicts the emitted photon spectrum of the star.  Thus a DA white dwarf can serve as the fundamental calibrator we seek---provided we can properly account for line-of-sight Galactic extinction. The HST CalSpec standards\cite{HST} that form the calibration basis for Hubble Space Telescope instruments also use this method.\cite{Bohlin1996} Figure \ref{fig:HST_WD} shows initial results obtained by Narayan {\it et al.},\cite{Narayan2015} who are attempting to establish a network of fainter DA calibration stars using photometry from HST in conjunction with ground-based spectroscopy. 

\begin{figure}[ph]
\centerline{\includegraphics[width=\textwidth]{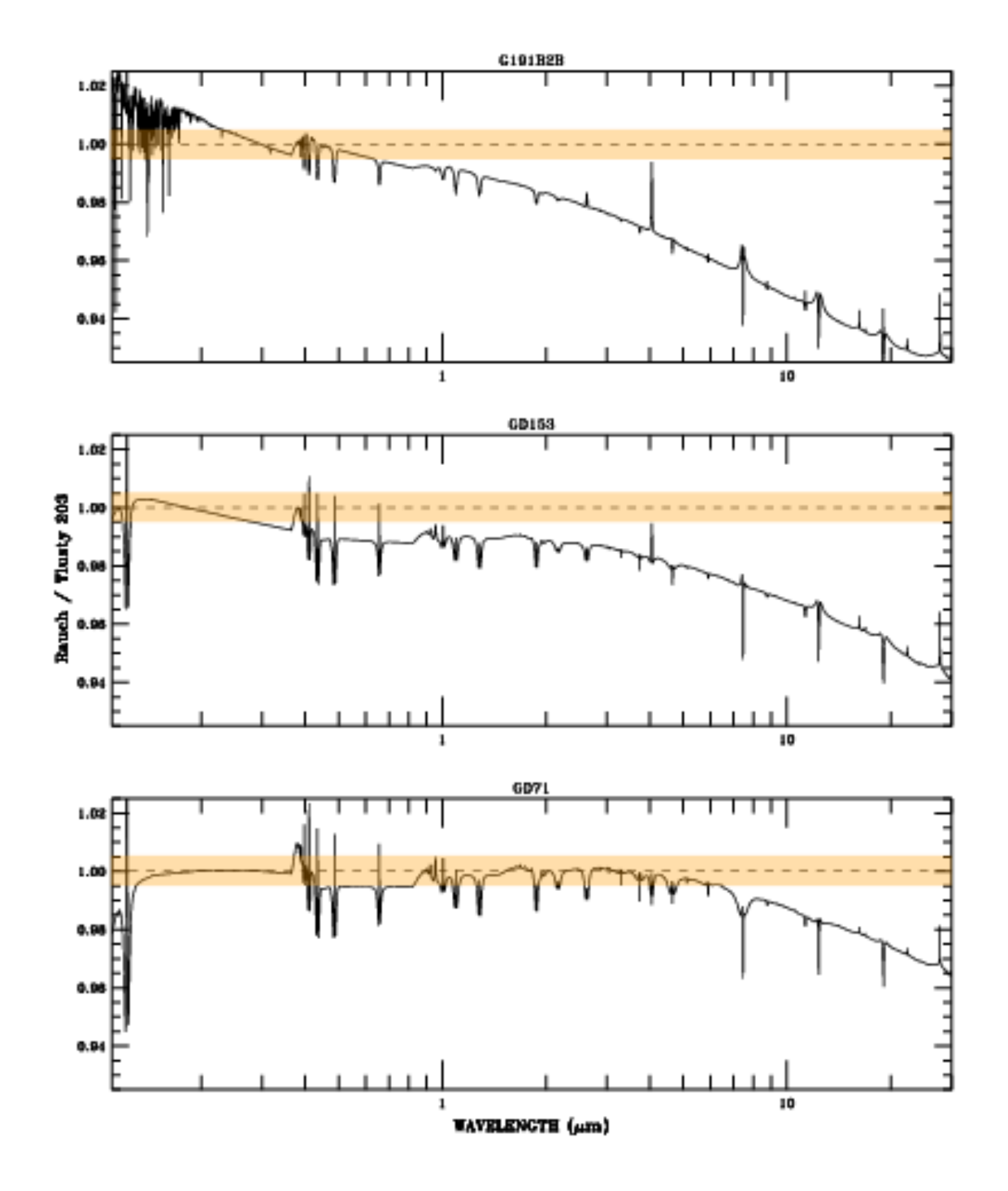}}
\vspace*{8pt}
\caption{The panels show the ratio of theoretically computed DA white dwarf spectra {\it vs.} wavelength, for two different state-of-the-art atmospheric modeling codes when the same values of $T_{\textnormal{eff}}$ and $\log(g)$ are used.  The orange bands correspond to 1\% uncertainties. The differences are attributed to different parameter choices for Stark broadening, different radiative transfer line lists, \textit{etc}. From R. C. Bohlin {\it et al.}\cite{Bohlin2014}, used with permission.
\protect
\label{fig:WDerrors}}
\end{figure}

Limitations to this method include the following:

\begin{enumerate}
\item Currently, the models produced by different groups exhibit discrepancies at greater than the 1\% level.\cite{Bohlin2014}. Figure \ref{fig:WDerrors} illustrates reported discrepancies between models that use the same values of $\log(g)$ and $T_{\textnormal{eff}}$.  This \emph{difference} between computed emission spectra is presumably a lower bound on the systematic uncertainties in the theoretical models.
 
\item Many of the DA standards currently in use are far too bright to be observed directly with survey instruments on large aperture telescopes. These objects are many magnitudes brighter than the saturation limit for the baseline 15-second exposure time on LSST, for example. Taking sub-second exposures for calibration purposes introduces new systematic errors due to shutter artifacts and atmospheric scintillation effects.
 
\item Any scattering by particulates along the line of sight (extinction, in the language of astronomy) distorts the spectrum, so that light arriving at the top of the atmosphere does not follow the theoretically computed emission spectrum.

\item Any source of temporal variability (rotation, convection, magnetic activity, {\it etc.}) in the stellar atmosphere is a potential source of systematic error. A sobering example of this effect was the recent realization \cite{BandL2015} that the commonly-used\cite{Sullivan2011} calibration star BD+17 has suffered a 4\% change in apparent magnitude over a six year interval.
\end{enumerate}

The spectral distortion due to extinction along the line of sight can be corrected for, given a prior assumption on how the scattering cross section varies with wavelength. If we fix the wavelength dependence of extinction to conform to one of the approximations commonly associated with dust in the Milky Way, such as that due to Fitzpatrick,\cite{Fitzpatrick1999} then the combined spectroscopic and broadband observational data can be analyzed as a three-parameter fit, with degrees of freedom that correspond to $T_\textnormal{{eff}}$, $\log(g)$, and the column density of dust along that particular sightline. Narayan {\it et al.}\cite{Narayan2015} are pursuing this approach, using a suite of DA stars that are faint enough to avoid saturation in the LSST catalog. 

If the discrepancies that presently exist between different modeling codes can be identified and addressed, then DA white dwarf stars could well serve as calibration standards at the sub-percent level of precision. The discrepancies are seen in the widths and (for some) the cores of the hydrogen absorption lines. Observations of these line shapes can be used to tune adjustable parameters in the modeling codes. Understanding the continuum spectral shape is critical to using these objects as celestial calibrators, however, and there are clear discrepancies in their overall spectral shapes. Both validation and verification of the modeling codes is required to establish that the theoretical predictions are reliable. Resolving the modeling discrepancies and establishing a reliable set of model spectra, in the context of a validation and verification campaign, strike us as well-motivated objectives. 

\subsection{Calibrated Detectors as a Flux Metrology Standard}

An alternative to using knowledge of the spectra of celestial sources is to use well-calibrated Silicon photodiodes as transfer standards for flux calibration. This approach relies on national metrology laboratories such as  NIST to provide detectors that are tied to fundamental SI flux metrology standards. Stubbs \& Tonry\cite{Stubbs2006} advocated breaking the flux calibration challenge into two distinct measurement problems: 1) explicit determination of atmospheric transmission, and 2) a detector-based calibration of instrumental response.  In the intervening decade the detector-based calibration method has been used to map out the response function of the CTIO 4m and MOSAIC imager\cite{CTIO}, PanSTARRS\cite{PSlasercal}, the CTIO 4m and the Dark Energy Camera,\cite{DECal} and the CFHT prime focus imager.\cite{SNLS2} The use of calibrated detectors is also a central element of the calibration plan for LSST,\cite{LSST} in conjunction with dedicating an auxiliary telescope to the determination of atmospheric transmission. 

The primary benefit of the detector-based method is that photon detection efficiency of the photodiode can be established with a precision of a few parts per thousand. This calibration method is the only one we know of where uncertainty in the calibration standard is low enough to permit a credible sub-percent determination of the instrumental response function.  

\begin{figure}[ph]
\centerline{\includegraphics[trim={0in 1in 0in 0.9in},clip, width=\textwidth]{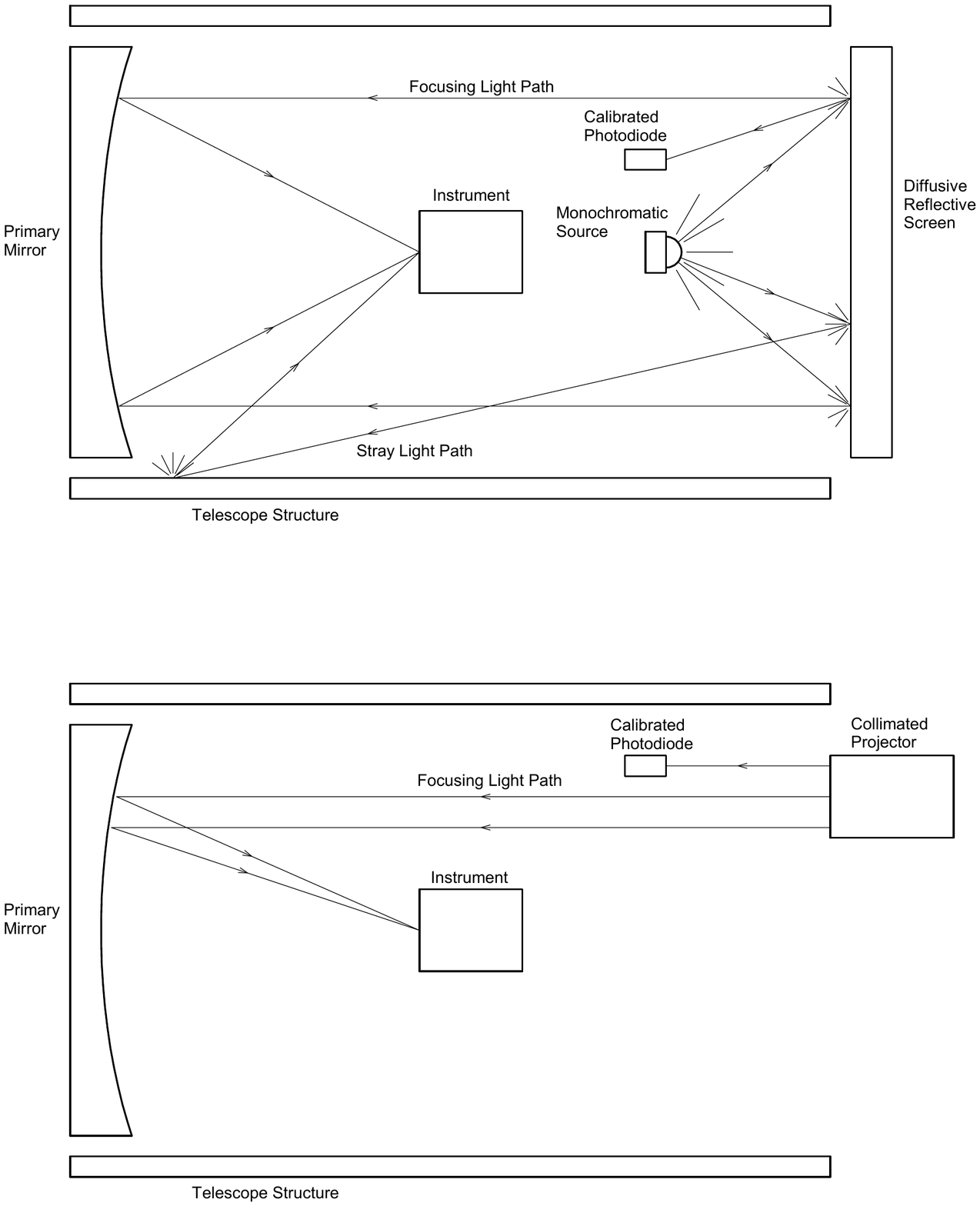}}
\vspace*{8pt}
\caption{Technique for tying instrument calibration to a standard Si photodiode.  Monochromatic laser light illuminates a flat-field screen at the entrance pupil of the telescope and a standard photodiode monitors the irradiance at the pupil. The upper panel shows light reflected from a flat-field screen (sample rays shown) in the telescope dome, and the lower panel shows a collimated beam from a projector. In the flat-field screen case the calibration illumination over-fills the phase space of angles incident on the pupil, and stray light paths produce systematic errors. In the collimated projector case the range of angles is limited to the field of view of the instrument but at the expense of under-filling the pupil position and ray angle distributions. In both cases the monitor photodiode is used to normalize the flux detected by the instrument. Iterating through wavelengths allows the instrumental response function to be tied to the known quantum efficiency of the diode. 
\protect
\label{fig:pdmethod}}
\end{figure}

The basic method is shown in Figure \ref{fig:pdmethod}. A narrowband illumination source injects photons into the pupil of the instrument.  This light is subject to all the loss mechanisms in the system---mirror reflection losses, air-glass interface losses, filter transmission function, and detector quantum efficiency---before ultimately exciting a signal in the detector. We need to establish the relative detection efficiency {\it vs.} wavelength for the rays that correspond to focusing light paths through the system, across the entire input pupil. The irradiance of the beam entering the pupil is monitored with a calibrated photodiode. To the extent that the irradiance of the illumination is independent of wavelength, mapping out the ratio of the number of photons detected in the instrument $N_{\textnormal{det}}$ (in a fixed exposure time) to the incident photon flux $N_0$ (as monitored by the photodiode) constitutes a measurement of the \emph{relative} instrumental response function, $R(\lambda)$, as a function of wavelength.  The instrument response function is 
\begin{equation}\label{IRF}
R(\lambda) = \frac{N_{\textnormal{det}}(\lambda)}{N_0(\lambda)} = \frac{N_{\textnormal{det}}(\lambda)}{I_P(\lambda)~ Q_P(\lambda)} ~,
\end{equation}
where $I_P(\lambda)$ and $Q_P(\lambda)$ are the photocurrent and quantum efficiency of the photodiode, respectively. The quantum efficiency of the phototdiode is known while the two unknown quantities on the right side of equation \ref{IRF} can be directly measured in the telescope dome---in principle providing a high-precision determination of $R(\lambda)$. 

\begin{figure}[ph]
\centerline{\includegraphics[width=\textwidth]{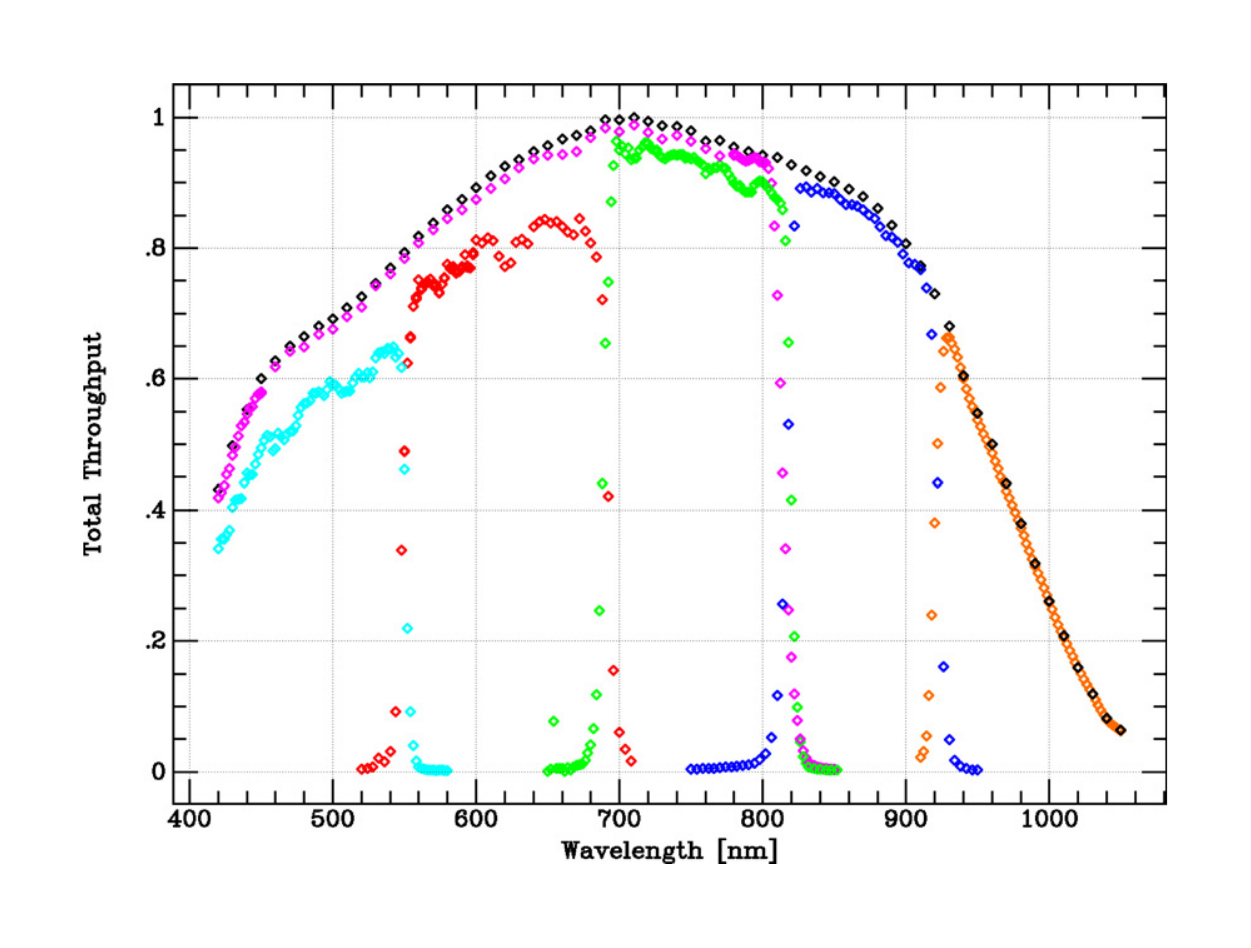}}
\vspace*{8pt}
\caption{Photodiode-based mapping of PanSTARRS system response function, from Stubbs {\it et al.}\cite{PSlasercal} (\copyright AAS. Reproduced with permission.) System throughput for various passbands is shown, \textit{vs.} wavelength. The central point of this plot is that the different filters are tied together onto a common flux scale, up to a single overall multiplicative factor.
\protect
\label{fig:laserscan}}
\end{figure}

An example of the data produced with this method is shown in Figure \ref{fig:laserscan}.

Although attractive in principle, this technique is non-trivial in practice and has yet to be fully implemented and demonstrated successfully.  Obstacles encountered with the method include the following: 

\begin{enumerate}

\item It is difficult to engineer a source of illumination in the dome that replicates a source at infinity, with collimated beam of uniform irradiance across the full input pupil of the telescope. 

\item Stray and scattered light paths from the in-dome light source produce detectable photons on the focal plane, but don't contribute to the detected flux from a celestial point source. This effect is a well-known gremlin in the flat-fielding of wide-field imaging instruments, and is responsible for the inconsistencies seen among dome flats, twilight sky flats, and the instrument's response to point sources.  

\item Multiple scattering light paths at wavelengths that are highly attenuated by interference filters (on the skirts of the filter response profile) is a particularly insidious source of systematic error. 

\end{enumerate}

An initial attempt to test for consistency compared\cite{Stubbs2012,Tonry2012} broadband PanSTARRS photometry of CalSpec spectrophotometric standards with a prediction based on the reported photon spectra of these objects, in conjunction with the system response function shown in Figure \ref{fig:laserscan} and an assumption (as opposed to a measurement) of atmospheric transmission.  The results they obtained are shown in Table \ref{tab:consistency}. 

\begin{table}[h]
\centering
\caption{PanSTAARS1 Photometric Consistency Checks.  The columns are the filter used, average difference between detected and predicted flux for standard stars, scatter among  $\sim 24$ observations, average difference between PS1 magnitude and SDSS magnitude, RMS scatter, and the number of stars compared.  From Tonry {\it et al.}\cite{Tonry2012}}
 {\begin{tabular}{rccccc}
\hline
Filter & Std & $\pm$ & SDSS & $\pm$ & N \\
\hline
$g_{P1}$ & $-$0.004 & 0.007 & 0.014 & 0.012 & 2644 \\
$r_{P1}$ & $-$0.005 & 0.006 & $-$0.019 & 0.010 & 3072 \\
$i_{P1}$ & 0.008 & 0.009 & 0.008 & 0.011 & 2850 \\
$z_{P1}$ & $-$0.009 & 0.007 & 0.015 & 0.011 & 2816 \\
$y_{P1}$ & 0.005 & 0.010 & 0.001 & 0.013 & 2150 \\
$w_{P1}$ & 0.002 & 0.011 & --- & --- & --- \\
\hline
\end{tabular}\label{tab:consistency} }
\end{table}

The Dark Energy Survey has undertaken a campaign to monitor the variable aspects of atmospheric transmission (aerosols, water vapor, ozone, and clouds) during survey observations, as well as the survey's instrumental response function.  Analysis of those data is under way.\cite{Burke2015}  Comparing the results obtained with this method to more traditional photometric calibration methods will be an important step towards realizing the goal of sub-percent precision. 

\section{Accounting for Attenuation by the Earth's Atmosphere, and by Galactic Extinction}

Regardless of what method is used for astronomical relative flux calibration, all extragalactic targets and most calibration stars inevitably lie both above the Earth's atmosphere and behind layers of interstellar dust in the Milky Way.  Achieving sub-percent knowledge of the cosmologically relevant spectral energy distributions of these sources requires that we account for the attenuation due to both the atmosphere and the interstellar medium.

The traditional definition of the luminosity of an astronomical object is the irradiance it produces at the top of the Earth's atmosphere.  This pragmatic definition accounts for atmospheric extinction, but ignores the effects of galactic dust, and consequently is inadequate for sub-percent photometry.  It is essential to account for both.  Fortunately, the precision with which we must determine the attenuation due to these effects scales inversely with their severity.  To achieve 1\% precision in attenuation correction for a process that knocks out only 10\% of the light, we need only measure the attenuation to a fractional precision of 10\%.

\subsection{Atmospheric Extinction}

There are two traditional methods used to correct for atmospheric extinction:  one is to assume that the attenuation obeys a Bouguer law so that the apparent magnitude of an observed object can be extrapolated to zero airmass as its zenith angle changes through the night; the other is to observe a calibration object at the same zenith angle and apply its observed attenuation to the object of interest.  Either of these methods is effective at the 10\% level of precision, but pushing the the sub-percent regime requires more care.

According to the Bouguer law, the flux $\phi$ of a beam of light passing through the atmosphere decreases exponentially with the airmass $\chi$ of the observation.  That is,
\begin{equation}\label{bouguer}
 \phi(\chi,\lambda) = \phi_0(\lambda) \; \mathrm{e}^{-\alpha(\lambda) \chi} ~,
\end{equation}
where $\phi_0$ is the sought-for flux of the beam at the top of the atmosphere, and $\alpha(\lambda)$ is a wavelength-dependent factor that accounts for the specific composition of the air.  The airmass $\chi$ is simply the secant of the zenith angle.  Thus $\log(\phi)$ (or the apparent magnitude) is linear in $\chi$ with intercept $\log[\phi_0(\lambda)]$.  By observing the object at various zenith angles through the night, the unattenuated (zero airmass) magnitude is easily extrapolated.

There are two main difficulties with this technique.  First is that it takes time, and during that time the characteristics of the atmosphere---as accounted for in $\alpha(\lambda)$---may well change.  Also, in the survey environment a source may not be monitored for long periods.  The second problem is that $\alpha$ changes from observation to observation, diluting the precision of comparisons from one object to another, or even for a particular object, from one observation to another.  Basically, for a given zenith angle, the optical depth---the product $\alpha \chi$---is not really known accurately.

Difficulties with the use of celestial calibration sources have already been discussed.  In order to account for atmospheric extinction at high precision, we must obtain a real-time evaluation of its transmission spectrum.

Fortunately, certain aspects of atmospheric attenuation are deterministic, depending almost entirely on the barometric pressure at the base of the atmosphere.  The basic chemical composition of atmospheric air is well known and quite stable, so it is possible to use our well-established understanding of atomic and molecular optics to model the propagation of light through it.  The MODTRAN atmospheric modeling computer code\cite{modtran} has been successful in performing this feat and is widely used in the remote sensing community.

Some aspects of the atmosphere are non-deterministic and time-varying, however.  There is weather, after all.  Non-deterministic atmospheric attenuation processes include
\begin{enumerate}
\item scattering from aerosols in the atmosphere, where the dependence of scattering cross section on wavelength is determined by both the shape and size distribution of the particles,

\item absorption by water vapor along the line of sight, which occurs at discrete wavelengths that are often saturated with an airmass dependence that does not obey the Bouguer law,

\item attenuation due to ozone in the upper atmosphere, and

\item scattering by ice and water particulates in clouds.

\end{enumerate}

Numerous researchers are developing methods of real-time direct measurement of atmospheric attenuation, including the use of LIDAR \cite{LIDAR}, differential GPS water vapor detection \cite{BlakeGPS}, and various balloon,\cite{Brown2013} rocket,\cite{ACCESS} and satellite-borne\cite{Albert2012} calibrated sources observed from the ground.  A full description of these efforts is beyond the scope of this paper, but we note that the LSST project has dedicated \cite{LSST} a 1.5 meter class telescope to this task.
\subsection{Galactic Extinction}

The precise measurement of the apparent magnitudes of extragalactic sources requires a correction for Galactic extinction. This correction clearly requires knowledge of the wavelength-dependent scattering cross section along any given line of sight. In this case we don't have the ability to modulate the line of sight through the attenuating medium, as can be done for atmospheric attenuation. The dust in the Milky Way scatters light at optical wavelengths, and emits blackbody radiation with an effective temperature of around 20 K. So we can use a combination of emission and attenuation data to constrain the amount of Galactic extinction along any line of sight. Early work\cite{SFD} used the IRAS infrared survey to produce an all-sky map of Galactic extinction using infrared emission. More recent 3D dust maps were produced using the observed perturbation of the apparent colors of stars.\cite{Green} Using the background density of galaxies to map out large scale structure in the Universe requires that we know how to properly correct for extinction in the Milky Way, and the determination of photometric redshifts similarly requires correcting the colors of galaxies for the spectral distortion (``reddening'') due to Milky Way dust.   
 
\section{Conclusions, and Suggested Next Steps}

Systematic errors from relative flux calibration prevent our making full use of increased catalogs of type Ia supernovae. Identifying and overcoming the multiple sources of systematic error seems entirely tractable, as evidenced by the ability of numerous projects to make differential astronomical flux measurements at precisions better than a part per thousand.  New techniques for characterizing the response of instruments, and careful corrections for atmospheric and Galactic attenuation are integral parts of this endeavor.  By pursuing a combination of techniques, including the use of DA white dwarf stars as celestial standards in conjunction with calibrated photodiodes as terrestrial standards, we can assess our progress in suppressing systematic errors as we enter the sub-percent regime.

The ``precision frontier"\cite{frontier} in astronomy is ripe for progress.  Ongoing efforts include the following.

\begin{itemize}

\item The ACCESS suborbital rocket payload will connect\cite{ACCESS} NIST photodiode calibration to bright stars. 

\item Wide-field photometric survey systems have installed progressively more sophisticated instrumentation both to monitor atmospheric transmission and to map out the response functions of their instruments. 

\item Combined analysis of Planck, Gaia, PanSTARRS, SkyMapper, and Dark Energy Survey data have led to improved constraints on Galactic extinction.

\item Wide field surveys such as DES, LSST, and WFIRST are pushing to implement improved precision for flux calibration. 

\end{itemize} 

We look forward to a decade of continual progress in achieving improved precision in relative astronomical flux calibration, with a commensurate improvement in our ability to constrain the nature of the dark energy.

\section*{Acknowledgments}
 
CWS is indebted to Nick Suntzeff for his patient explanation of the nuances of astronomical flux calibration during our work together on the ESSENCE supernova project.
We are also grateful to Justin Albert, Tim Axelrod, Ralph Bohlin, David Burke, Chuck Claver, Claire Cramer, Jim Gunn,  Doug Finkbeiner, Patrick Ingraham, Zeljko Ivezic, Robert Lupton, Keith Lykke, Gene Magnier, Jennifer Marshall, John McGraw, Gautham Narayan, Armin Rest, Abi Saha, John Tonry, Tony Tyson, and Peter Zimmer for informative conversations on this topic. 

This work was supported by the LSST Project, NSF under grant AST=0507475, and the US Department of Energy under grant SC0007881.


\end{document}